%% file: 0_main.tex
\documentclass[sigconf]{acmart} 
\usepackage[utf8]{inputenc}
\DeclareUnicodeCharacter{202F}{\,}
\usepackage{booktabs} 
\usepackage{geometry}
\usepackage{threeparttable}
\usepackage{array}

\AtBeginDocument{%
  }

\begin{document}

\title[Diverging Expectations for Enterprise AI as Generative AI Took Shape]{Beyond the Personal Assistant: How Expectations for Enterprise AI in Teamwork Diverged as Generative AI Took Shape, 2023–2025}

\author{Qing Xiao}
\affiliation{
  \institution{Human-Computer Interaction Institute, Carnegie Mellon University}
  \city{Pittsburgh}
  \state{Pennsylvania}
  \country{USA}
}
\email{qingx@cs.cmu.edu}

\author{Xinlan Emily Hu}
\affiliation{
  \institution{Institute for Data Systems and Society, Massachusetts Institute of Technology}
  \city{Cambridge}
  \state{Massachusett}
  \country{USA}
}
\email{xehu@mit.edu}

\author{Mark E. Whiting}
\affiliation{
  \institution{Pareto \& Computational Social Science Lab, University of Pennsylvania}
  \city{San Francisco}
  \state{California}
  \country{USA}
}
\email{mark@pareto.ai}

\author{Arvind Karunakaran}
\affiliation{
  \institution{Department of Management Science and Engineering, Stanford University}
  \city{Palo Alto}
  \state{California}
  \country{USA}
}
\email{arvindka@stanford.edu}

    \author{Hong Shen}
\affiliation{
  \institution{Human-Computer Interaction Institute, Carnegie Mellon University}
  \city{Pittsburgh}
  \state{Pennsylvania}
  \country{USA}
}
\email{hongs@cs.cmu.edu}

\author{Hancheng Cao}
\affiliation{
  \institution{Goizueta Business School, Emory University}
  \city{Atlanta}
  \state{Georgia}
  \country{USA}
}
\email{hancheng.cao@emory.edu}

\begin{abstract}
%HCI often draws on users’ articulated needs and expectations to explore design opportunities for emerging technologies. Yet, these accounts are shaped by how technologies take form over time. We examine this dynamic through a two-phase interview study of what workers expected enterprise AI to do for their teams in a project-based software development organization. We interviewed 15 practitioners in 2023 and re-interviewed 10 in 2025. In 2023, participants expected enterprise AI built around teams: systems integrating signals across workplace tools, detecting coordination problems, and intervening in team processes. By 2025, no such team-focused systems had entered their work, while personal AI assistants were prevalent. Participants’ positions then diverged: some continued to expect team-oriented AI, some rejected it as undesirable or infeasible, and some no longer articulated it at all. These trajectories show how design possibilities can remain open-ended, narrow, or disappear as technologies take shape, with implications for how HCI elicits, interprets, and designs from user expectations over time. 
HCI often draws on users’ articulated needs and expectations to explore design opportunities for emerging technologies. Yet, these accounts are shaped by how technologies take form over time. We examine this dynamic through a two-phase interview study of enterprise AI in a project-based software development organization. We interviewed 15 practitioners in 2023 and re-interviewed 10 in 2025. In 2023, participants expected enterprise AI built around teams: systems integrating signals across workplace tools, detecting coordination problems, and intervening in team processes. By 2025, no such team-focused systems had entered their work, while personal AI assistants were prevalent. These expectations diverged from one another: some expectations were deferred, some rejected as undesirable or infeasible, and some no longer articulated as expectations they had held. These trajectories show how design possibilities can remain open-ended, narrow, or disappear as technologies take shape, with implications for designing enterprise AI and interpreting present-day user accounts. 
\end{abstract}




\maketitle

\input{1_intro}

\input{2_related}
\input{3_method}
\input{4_phase1}
\input{5_phase2}
\input{6_discussion}
\input{7.limitations}
\input{8_conclusion}

\bibliographystyle{ACM-Reference-Format}
\bibliography{sample-base}

\appendix

\end{document}

%% file: 1_intro.tex
\section{Introduction}
\label{sec:intro}

HCI often draws on users’ articulated needs and expectations to explore design opportunities for emerging technologies~\cite{spath2012user,lindgaard2006user,bannon2012design,sengers2021speculation}. These accounts help researchers surface problems, imagine possible systems, and reason about what may be worth building. In many design processes, they are elicited early and inform subsequent ideation and prototyping~\cite{designcouncil_double_diamond,karapanos2009user}. Yet, such accounts are necessarily situated in time. Expectations are formed in relation to what people know a technology to be, what they have experienced it doing, and what possibilities they can imagine from that vantage point. As technologies themselves evolve, the accounts researchers collect from users may change with them~\cite{borup2006sociology,vanlente2012navigating,brown2003sociology}.

This temporal problem is especially important for fast-changing technologies, where the capabilities, interfaces, and uses that shape users’ expectations can shift substantially over a short period. Enterprise AI provides a particularly useful case. We use \emph{enterprise AI} broadly to refer to AI incorporated into organizational work and workflows~\cite{mohammed2026enterprise,davenport2023all}. Since the arrival of generative AI in late 2022, workers have encountered rapidly changing forms of AI at work. Personal AI assistants such as ChatGPT and Copilot have become common tools for writing, coding, and analysis~\cite{salari2025impacts,shao2025future,brachman2024knowledge,xiao2025might}, while research and industry have continued to explore systems that operate across shared workplace tools, maintain context about collective activity, and support coordination among team members~\cite{dell2025cybernetic,Spataro2025FrontierFirm,SuperAGI2025TeamComm}. Enterprise AI therefore offers a setting in which both the technology and the possibilities workers associate with it can change quickly.

We focus on one part of this design space: what workers expect AI to do for their teams. Teamwork is a consequential setting for such expectations because work in organizations is deeply interdependent. Decades of CSCW research document persistent problems of coordination, awareness, and team viability~\cite{cao2021large,whiting2019did,hinds2003out}. Recent studies find that generative AI can concentrate effort on individual tasks and productivity while shifting work away from collaborative activities~\cite{hoffmann2024generative,song2024impact}. CSCW has long emphasized that collaboration depends on situated social and organizational work that technological systems do not simply supply a priori~\cite{andriessen2012working,ngwenyama1997groupware,xiao2026can}. These tensions make AI for teamwork a useful strategic research setting for asking a broader HCI question: \textit{what becomes of design possibilities articulated while an emerging technology is still taking shape?}

A single round of elicitation cannot answer this question. It can tell us what people articulate at one moment, but not how that account relates to the possibilities they articulated before. An expectation about a technology that is present today may never have existed, or may have been reconsidered. Existing longitudinal HCI research shows the value of returning to the same people as their experiences and practices change~\cite{jungselius2025tracing,blaising2021making}. Much less work has followed the same articulated expectations of an emerging technology across a period in which the technology itself is evolving and changing at the same time. Doing so requires more than asking users retrospectively what they once wanted: it requires an earlier record against which later accounts can be compared.

We study such a record in a distributed, project-based software development organization. Contributors are geographically dispersed, frequently part time, and staffed to individual client projects for their duration rather than to standing teams. Coordination among people who may not have worked together before is therefore a recurring requirement rather than an occasional one, a structure increasingly common across knowledge work as project-based organizing, platforms, and on-demand team assembly have spread~\cite{sydow2004project,valentine2025flash,vallas2020platforms,gorton1996issues}. The organization also adopted AI early, deploying GitHub Copilot and GPT-based assistants internally in early 2023. This setting allowed us to record expectations at one moment in the emergence of generative AI and return two years later after participants had accumulated substantially more experience with AI in their everyday work.

We interviewed 15 members of the organization in February and March 2023 and re-interviewed 10 of them in February and March 2025. In both phases, we asked what participants expected AI to do for their teams. This repeated question makes the two accounts directly comparable and lets us examine change in what the same people articulated rather than differences between populations.\footnote{We use AI in the sense participants used it. In early 2023 their accounts moved between predictive analytics, systems that would act on a team's behalf, and the conversational assistants then newly available, and we retain that breadth because narrowing the term after the fact would remove the object of study. Throughout the paper, "team-level expectations" refers to what participants expected AI to do for a team as a unit, as distinct from what they expected it to do for themselves individually.} We organize the study around two questions:
\begin{itemize}
\item \textbf{Phase 1 (2023):} What did participants expect AI to do for their teams, and under what conditions did they consider such systems acceptable?
\item \textbf{Phase 2 (2025):} What became of those expectations two years later?
\end{itemize}

In 2023, participants articulated detailed and largely convergent expectations for AI in teamwork. They described systems that could read across workplace tools such as Jira, Slack, and GitHub; learn what normal looked like for a team; detect emerging coordination problems; combine signals that no single tool held; and intervene in team processes. They also placed limits on such systems: assessments should reach the person concerned before a manager, data collection should remain visible, and AI should inform rather than replace human judgment. Several participants explicitly distinguished these expectations from the conversational assistants then becoming available \textbf{(Phase 1)}.

By 2025, no system matching these expectations had entered participants’ work, while personal AI assistants had become routine. Their earlier expectations did not move together. Four participants \textit{deferred} them, continuing to hold and often further specify what they had wanted. Three \textit{rejected} them, remembering the earlier expectation but now viewing such systems as infeasible, misdirected, or undesirable. Three \textit{dropped} them: expectations they had articulated in 2023 no longer appeared in their 2025 accounts. These trajectories describe changes in participants’ articulated expectations rather than underlying beliefs we can directly observe; in particular, an absent expectation may have been forgotten, abandoned, or no longer treated as relevant \textbf{(Phase 2)}.

These divergent trajectories reveal something that a present-day user account alone cannot show: different histories can lead to superficially similar positions about what should be designed. For HCI, the implication is not simply that expectations should be elicited repeatedly. Rather, earlier and later accounts provide different kinds of design evidence. Deferred expectations preserve possibilities that remain open; rejected expectations reveal constraints and concerns that emerged with experience; and dropped expectations identify alternatives that later elicitation may no longer surface at all. Following expectations over time therefore provides a record not only of what users currently want, but of how the design space they articulate changes as technologies take shape. We use this record to consider implications for designing enterprise AI and, more broadly, for interpreting present-day user accounts in rapidly evolving technological domains.

%% file: 2_related.tex
\section{Background and Related Work}
\label{sec:related}

\subsection{AI at Work: Individual Use in Interdependent Settings}
\label{workplace}

Much recent work on AI in the workplace has focused on how individuals use and respond to AI. Studies examine automation of work tasks~\cite{uren2023technology,tschang2021artificial,mcelheran2024ai}, productivity gains~\cite{ju2025collaborating,weisz2025examining}, and support for decision-making~\cite{booyse2024barriers,vasconcelos2023explanations,deseriis2023reducing}. Research on technology adoption similarly identifies factors that shape whether workers use new systems, including perceived usefulness and ease of use~\cite{davis1989technology,marangunic2015technology}, as well as trust, risk, and transparency in AI~\cite{sohn2020technology,ma2025exploring,xu2022technology}. This work is important for understanding how AI becomes part of everyday work.

At the same time, individual AI use takes place within interdependent organizations. AI-generated outputs are shared with colleagues, incorporated into joint workflows, and can change how work is divided across roles. Research has therefore increasingly examined AI in relation to organizational culture, hierarchy, and workplace relationships~\cite{das2024sensible,kang2022stories,anthony2023collaborating,herrmann2023keeping}, alongside work on trust and appropriation in human--AI interaction~\cite{duan2024understanding,duan2025trusting,duan2025understanding}. Recent studies also show that individual AI use can have collective effects. Hoffmann et al. find that GitHub Copilot shifts effort away from collaborative project management and toward individual coding work~\cite{hoffmann2024generative}; Song et al. report changes in coordination time~\cite{song2024impact}; and Xiao et al. find that generative AI use in newsrooms can remain individualized and disconnected from collaborative workflows~\cite{xiao2026can}.

Our study builds on this work by focusing on what workers expected AI to do for their teams. Rather than treating individual and team-level uses as competing perspectives, we examine how expectations about AI for teamwork developed alongside workers' growing experience with AI in their own work.

\subsection{AI for Teamwork and Coordination}
\label{tech}

Supporting teamwork is a longstanding concern in CSCW. Distributed teams need to maintain awareness of who is doing what and coordinate work across people, tools, and time~\cite{gutwin2004group,Olson2009GroupwareCSCW,Baecker1993ReadingsGroupwareCSCW,Wallace2017TechnologiesMethodsValuesCSCW}. CSCW systems have supported this work by making activity visible through shared whiteboards~\cite{Prante2002DevelopingCSCWToolsIdeaFinding,Aytes1995CollaborativeDrawingTools,Pekkola2003DesignedUnanticipatedUse,gronbaek2024blended}, version-control platforms~\cite{Tsay2014LetsTalkGitHub,Dabbish2012SocialCodingGitHub}, and collaborative editors~\cite{Yeh2024UpdateIntervalRevealMethodSharedEditors,Wang2015DocuViz}. In distributed software development, issue trackers, CI/CD systems, and communication tools similarly provide infrastructure for coordination~\cite{espinosa2007team,Bjorn2024AchievingSymmetryHybridWork,hu2022distance}.

These systems help, but collaboration remains difficult. Remote and hybrid teams continue to experience problems with coordination, communication, and shared awareness~\cite{garro2021virtual,yang2022effects,strode2022teamwork,zhang2020web,cao2021large}, motivating continued work on awareness and information sharing~\cite{Meyer2025BetterBalancing,Liu2024LetInformationFlowAwareness,Wang2024MeetingBridges}. CSCW has also emphasized that effective collaboration depends on social and organizational practices that cannot be reduced to information exchange alone \cite{bjorn2014does,gutwin1998design,grudin1988cscw}.

Recent AI systems extend this line of work. Research prototypes use large language models to facilitate meetings, support discussion, and scaffold shared sense-making~\cite{kim2025beyond,zhang2025ladica,morrison2024ai}. Deployed systems transcribe and summarize meetings through tools such as tl;dv~\cite{Lomas2022TLDV} and Otter.ai~\cite{Schultz2019OtterMeetingAssistant}, organize project information through Notion AI~\cite{OpenAI2025NotionWorkflow} and SlackGPT~\cite{Keenan2023FutureOfWorkSlackGPT}, and support task completion through systems such as MoveWorks~\cite{Moveworks2025AIAgentMarketplace}. Team assistants and agents have also begun to appear~\cite{Spataro2025FrontierFirm,SuperAGI2025TeamComm}. At the same time, AI increasingly participates directly in the work that teams coordinate, raising questions about accountability, shared mental models, expertise, and the division of labor between humans and AI~\cite{Procter2023HoldingAIToAccountHealthcare,Andrews2023SharedMentalModelsHumanAITeams,rahe2025programming,li2024user,lewis2025generative}.

This literature shows that AI can enter teamwork in many ways: by supporting awareness, coordinating information, facilitating interaction, or contributing directly to task execution. Our study complements this system- and use-centered work by examining what practitioners themselves expected AI to do for their teams, and how those expectations changed over time.

\subsection{User Expectations and Design Opportunities Over Time}
\label{social}

HCI frequently draws on what users say they need, want, or expect when exploring technologies that do not yet exist. Needs assessment and requirements elicitation rely on users' articulated needs to inform design~\cite{spath2012user,lindgaard2006user}. Design fiction, speculative design, and scenario-based methods introduce possible futures to elicit reactions beyond current products~\cite{ringfort2025quality,rapp2019design,sengers2021speculation}. Participatory and co-design approaches similarly involve users in imagining alternatives and shaping design directions~\cite{bannon2012design,yu2025participatory,steen2013co}. These approaches do not assume that users can predict the right product. Instead, they leverage what users articulate to identify problems, values, and possibilities worth exploring.

HCI has also long recognized that what users say depends on what they have experienced. People can find it difficult to articulate needs for unfamiliar technologies~\cite{norman2013design,patnaik1999needfinding}, which has motivated probes and speculative methods designed to broaden what can be discussed~\cite{boehner2007hci,sengers2021speculation}. Longitudinal HCI research has examined how experiences with technologies change after adoption~\cite{karapanos2009user,kujala2011ux}. However, less attention has been paid to what happens to expectations that were articulated \emph{before} or during the emergence of a new technology, once users have had substantially different experiences with it.

Research on expectations and imaginaries suggests why this matters. Literature on the sociology of expectations shows that expectations change as technologies develop and as new possibilities become salient~\cite{borup2006sociology,vanlente2012navigating}. Expectations can be revised, sustained, or fully rejected over time~\cite{borup2006sociology}. HCI and CSCW research on sociotechnical imaginaries likewise shows that visions of technological futures are shaped by social context, professional identity, and experience~\cite{Jasanoff2009ContainingAtom,wang2023ai,zhong2025ai,kang2022stories}.

Most of this work examines expectations at one point in time or follows broader public and institutional discourse~\cite{borup2006sociology,Jasanoff2009ContainingAtom}. A smaller body of longitudinal qualitative work shows the value of returning to the same participants to understand changes that cannot be recovered from a single interview~\cite{jungselius2025tracing,blaising2021making}. We build on this latter approach by asking the same practitioners what they expected AI to do for their teams in 2023 and again in 2025. This allows us to examine not only what people expected at either point but also what became of expectations they had previously articulated as the technologies they encountered at work changed around them.

%% file: 3_method.tex
\section{Method}
\label{sec:method}

This study was conducted in two interview phases. We interviewed 15 members of a distributed, project-based software development organization in February and March 2023 (Phase 1), and re-interviewed 10 of them in February and March 2025 (Phase 2), for 25 interviews in total. We focus on depth within one organization: the same people, in the same setting, at the moment generative AI was first entering their work and two years later, when it had become routine. We analyze participants' expectations of AI for teamwork and what became of them over the two years between the phases. Both phases asked participants the same question about what they expected AI to do for their teams, which is what makes the comparison possible. 

\begin{table*}[ht]
\centering
\begin{threeparttable}
\caption{Study design across the two phases.}
\label{tab:design}
\begin{tabular}{@{}p{1.6cm}p{3.2cm}p{4.2cm}p{4.2cm}@{}}
\toprule
\textbf{Phase} & \textbf{Period} & \textbf{Interview protocol} & \textbf{AI in participants' work} \\
\midrule
Phase 1 & February and March 2023, months after the public release of ChatGPT. 15 participants (P1--P15). &
(1) Routine practice of distributed teamwork and its frictions. (2) Familiarity with AI tools, not restricted to language models. (3) \textit{Expectations towards AI}: what AI should do for their teams, and under what conditions. &
Predictive analytics and reporting tools already in use. GitHub Copilot and GPT-based coding assistants newly deployed internally. Generative tools tried by some participants, not yet routine. \\
\addlinespace
Phase 2 & February and March 2025. 10 of the original 15 (P2, P4, P5, P6, P9--P14). &
(1) Current AI use across coding, documentation, communication, project management, decision support. (2) Effects on own and colleagues' contribution. (3) Team norms around pace, disclosure, accountability. (4) \textit{Expectations towards AI}, retrospectively and going forward. &
ChatGPT, Claude, and Copilot in routine individual use. Meeting transcription tools adopted. No system reading across Jira, Slack, GitHub, or project plans at the team level. \\
\bottomrule
\end{tabular}
\end{threeparttable}
\end{table*}

Following two-phase interview designs in HCI and CSCW~\cite{blaising2021making,jungselius2025tracing}, we treat the second phase as a return to the same people rather than as a separate study. We do not claim continuous observation of the intervening period. What we observe is two accounts from the same person, given two years apart, and what changed between them. \autoref{tab:design} summarizes the design, including what AI was available in participants' work at each phase.

Two definitional choices follow from this object of analysis. First, we use the term AI "emically" i.e., in the sense participants interpreted and used it \cite{jorion1983emic}. In 2023 our participants did not treat AI as a single product category: their accounts moved between predictive analytics, machine learning over the organization's project history, systems that would act on a team's behalf, and the conversational assistants then newly available. We keep this breadth, since expectations form before the technologies that might satisfy them exist, and narrowing the term to generative AI afterwards would leave out much of what we are studying. Second, we distinguish throughout between \textit{team-level expectations}, meaning what individual participants expected enterprise AI to do for a team as a unit, and expectations of what AI would do for them individually as personal assistants. This study focuses on the team-level expectations articulated by the individual members. We report individual expectations only where they bear on the comparison, which they do because they rose over the same period in which the team-level ones did not. \autoref{tab:design} records which forms of enterprise AI were present in participants' work at each phase, so that the reader can see what each round of expectations was formed against.

\subsection{Study Site and Participants}
\label{sec:context}

The organization assembles remote teams to deliver enterprise software and digital transformation projects. Contributors are geographically distributed and staffed to a project for its duration rather than to a standing team, so they coordinate with collaborators they may not have worked with before. Teamwork in this setting is both essential to delivery and difficult to sustain, which makes expectations of tools that might support it unusually explicit.

The site suits our question for a second reason. The organization and its workforce adopted AI tools early, beginning internal deployment of GitHub Copilot and GPT-based coding assistants in early 2023. We chose this organization because it is a good place to see the phenomenon. Its members were forming expectations about AI at the moment the technology arrived, which is exactly the process we wanted to observe. Choosing a case for what it makes visible follows standard practice in case study research~\cite{flyvbjerg2012five}.

\autoref{tab:participants} lists participants. They span software engineering, management, design, and consulting, with 3 to 35 years of professional experience (mean approximately 16). Their projects relied on distributed coordination tooling for both synchronous and asynchronous work, including chat, shared documentation, version control, ticketing, and project reporting.

Recruitment in 2023 proceeded through internal referrals facilitated by company liaisons and subsequent snowball sampling. All interviews were conducted with approval from the university's IRB and from the participating organization, and all participants gave informed consent. Participants are identified by number, and the organization is not named.

Ten of the 15 participants took part in Phase 2. The five who did not (P1, P3, P7, P8, P15) either did not respond to our invitations or could not be scheduled during the data collection period. Attrition was not concentrated in a role or seniority band: the five comprise two software engineers, two project managers, and one product manager, with 4 to 30 years of experience. In 2023, all five had articulated team-level expectations coded to the functions reported in \autoref{sec:phase1}. We cannot rule out differences on dimensions we did not measure. In this study, we report Phase 1 findings for all 15 and restrict cross-phase claims to the 10.

\begin{table*}[t]
\centering
\begin{threeparttable}
\caption{Participants’ Demographics.}
\label{tab:participants}
\Description{A five-column table listing the fifteen participants, P1 to P15, with their primary role, years of professional experience and the domain of that experience, areas of expertise, and whether they took part in the 2025 follow-up. Roles span software engineering, architecture, product and project management, consulting, and UX design. Ten participants took part in the follow-up and five did not.}
\begin{tabular}{@{}lllll@{}}
\toprule
\textbf{ID} & \textbf{Primary Role} & \textbf{Years of Experience} & \textbf{Expertise} & \textbf{2025 Follow-up} \\
\midrule
P1  & Software Engineer      & 10 (software)      & Full-stack development; accessibility \& UX & No \\
P2  & Technical Architect    & 3 (software)       & Software development; team communication & Yes \\
P3  & Software Engineer      & 30 (software)      & Software design; consulting; game design & No \\
P4  & Product Manager        & 35 (management)    & Game development; technical consulting & Yes \\
P5  & Technical Consultant   & 25 (software)      & Health tech; quantified-self; mobile health apps & Yes \\
P6  & Product Manager        & 15 (management)    & Browser development; team management & Yes \\
P7  & Project Manager        & 10 (management)    & Project \& client management & No \\
P8  & Project Manager        & 4 (management)     & Project coordination & No \\
P9  & Software Engineer      & 12 (software)      & Blockchain; technical architecture & Yes \\
P10 & Software Architect     & 20 (software)      & Digital platform services & Yes \\
P11 & UX Designer            & 8 (design)         & Real estate app design & Yes \\
P12 & Consultant / PM        & 30 (software)      & Full-stack development & Yes \\
P13 & Software Engineer      & 20 (software)      & Robotics \& automation development & Yes \\
P14 & Consultant             & 20 (software)      & Enterprise architecture & Yes \\
P15 & Product Manager        & 4 (management)     & Project management & No \\
\bottomrule
\end{tabular}
\end{threeparttable}
\end{table*}

\subsection{Data Collection: Two Interview Phases}

Phase 1 took place months after the public release of ChatGPT. Interviews were conducted remotely, lasted 60 to 90 minutes, and were held in English, audio recorded with consent, professionally transcribed, and pseudonymized. The protocol had three parts, each addressed to a component of the research question. We first asked participants to describe their routine practice of distributed teamwork and the frictions they encountered in it, in order to establish the problems against which any expectation of AI would be formed. We then asked about their familiarity with AI tools, deliberately without restricting the term to large language models, in order to record the range of systems they had in mind. We finally asked what they expected AI to do for their teams: which parts of teamwork they considered automatable or augmentable, which they considered to require human judgment, and under what conditions they would accept a system intervening. This third block was coded as \textit{expectations towards AI} and is present for every participant in the phase. 

Phase 2 took place two years later, by which point generative AI tools had diffused widely in participants' work. Nine interviews were conducted remotely under the same conditions as Phase 1. One participant (P13) asked to respond asynchronously in writing and was given the same protocol; the responses were returned over several days and analyzed alongside the others. The protocol had four parts. We asked how participants and their teams were using AI in practice, across coding, documentation, communication, project management, and decision support. We asked how AI had affected their sense of their own contribution and their view of their colleagues' contributions. We asked whether norms within their teams had changed, including around pace, disclosure, and accountability. Finally, and as in Phase 1, we asked what they expected AI to do for their teams, both retrospectively and going forward. This block was again coded as \textit{expectations towards AI} and is present for every participant in the phase. 

Two features of our research design address the risk that any difference between the phases is an artifact of how we asked. First, the expectations question was asked in both phases, so the comparison rests on the same prompt rather than on speculation in one phase and reported experience in the other. Second, before inviting comparison we reminded participants of what they had said in 2023, which gave them the opportunity to affirm, revise, or disown their earlier accounts. Several did revise them, and those revisions are themselves part of what we report.

\subsection{Analysis}

Our analysis proceeded in two stages, following two-phase qualitative designs that generate themes within each timepoint before linking across them~\cite{blaising2021making,jungselius2025tracing}. We first analyzed each phase on its own using reflexive thematic analysis~\cite{braun2019reflecting}. Members of the research team open-coded the transcripts from each phase independently, then met to consolidate their codes into a codebook for that phase, which was applied to all transcripts from that phase. We built the 2025 codebook without reference to the 2023 one so that categories specific to the later period could emerge rather than being read through the earlier ones. Disagreements were resolved through discussion within the research team rather than by calculating inter-coder reliability score or other agreement statistics. This is consistent with the reflexive approach we adopt, where coding is treated as interpretive work rather than as measurement.

We then linked the two phases. For each of the 10 returning participants, we placed the 2023 and 2025 transcripts side by side and compared their accounts within each code, asking what was retained, what was revised, what was newly present, and what was absent. One feature of this comparison needs stating. Our object is participants’ expectations for what enterprise AI could do for their teams; changes in the forms of AI they encountered between phases provide the technological context in which those expectations were re-articulated. In 2023 the form participants expected was built around teams; the form that had arrived by 2025 was personal assistants. The expectations question was the same in both phases and did not name either form, so when participants answered in 2025 in terms of the personal assistants they now used rather than the team-facing systems they had described in 2023, we treated that shift in reference point as data rather than as a mismatch between the phases. The three positions we report in \autoref{sec:phase2}, deferred, rejected, and dropped, were derived from this paired comparison. They are our categories rather than distinctions participants drew, and we say so in the text where they are introduced.

Two limits follow from this research design. First, both phases consist of accounts given in interviews. Where we report change between 2023 and 2025, we report change in how participants articulated their expectations, and we do not claim direct access to what they believed at either point. Second, where a 2023 expectation is absent from a participant's 2025 account, we cannot determine from interview data whether it had been forgotten, was no longer endorsed, or was not treated as relevant to the question asked. We report the absence and the reasons participants gave, and we mark this limit where it matters to the claim.

%% file: 4_phase1.tex
\section{Findings: Expectations of Enterprise AI Built Around Teams in 2023 (Phase 1)}
\label{sec:phase1}

In 2023, participants held detailed and largely convergent expectations of what AI would do for their teams. They named the data their teams already generated, the signals future AI should extract from it, the point at which it should act, and the limits it should observe. Together, they expected enterprise AI to sit as a layer over the tools they already used: a system that would read tickets in Jira, messages in Slack, commits in GitHub, and dates in project plans, and convert them into early information about how a team was doing.

\subsection{The Problems These Expectations Were Anchored To}
\label{ex1}

Participants attached their team-level expectations to two problems they described as chronic in distributed, project-based software work. The first was that underperformance became visible only after it had caused damage. Contributors worked part time, across time zones, and only for the length of a single project, which in their accounts meant that disengagement produced no immediate signal. P2 described a sprint in which a developer produced nothing until deadlines were already at risk: \textit{“We were relying on some developers, and one of them in particular just had done nothing for an entire sprint block plus. And it wasn't until I just saw what he had.”} P5 described a slower version of the same problem: \textit{“We've had some critical people and things are great, and then all of a sudden they just sort of drop off, and they're not communicating as much or not delivering as much.”} In both accounts, detection depended on whether someone happened to be paying attention. P3 considered this an unreasonable basis for managing a project: \textit{“You're kind of asking too much of the team manager who's making the judgment to say whether they're gonna guess the project will succeed, especially when it's just kind of a human feeling.”}

The second problem was that structured communication did not produce shared understanding. Jira and Slack, participants said, organized information without conveying who was doing what, or how they were doing. P7 described the resulting gap: \textit{“Slack is our heartbeat, but sometimes I don't even know what some people are doing. I just know I haven't heard from them.”} P6 attributed the frictions that followed to unclear expectations and incomplete handoffs rather than to personality, and noted that they were worst when contributors left a project without transferring what they knew.

Participants said existing mechanisms addressed neither problem. Status updates were irregular, and peer feedback arrived only after the work was finished and only when someone chose to give it. As P1 noted, \textit{“It's completely optional. So if your team never gives feedback, you'll never get any.”} P2 said the same of the surveys the organization ran: \textit{“None of that survey stuff had anything to do with the identification of the problem, or the resolution of it.”} Participants described the gap as being filled by individuals: experienced contributors took on late coordination and repair work, which they said kept projects delivering while concentrating the cost on a few people.

\subsection{What Participants Expected AI to Do in Future Teamwork}

We group these accounts into three functions. Sensing, diagnosis, and mediation are our categories rather than a design any participant laid out. Most described more than one of the three.

\subsubsection{Sensing: Expecting AI to Detect When a Team Departs from Its Own Baseline}
\label{ex2}

Participants expected AI to read the records their work already produced and report when a team deviated from its own baseline. The baseline was to be learned from the team's history rather than set in advance, which in their accounts distinguished this from the reporting tools they already had.

Silence was the signal participants named most often, and they described it as the one they most often missed. P6 wanted a system that would formalize an instinct he already used: \textit{“My Spidey sense, if it's too quiet, something's wrong.”} P5 described the same idea as a comparison against normal volume: \textit{“What is the volume compared to what would maybe be considered normal? And then is there a sort of downtrend?”}

Other participants located the signal in work artifacts. P6 wanted to track whether tickets moved at all: \textit{“If you have 100 tickets and you're in the first milestone, is there movement or activity on the tickets from lane to lane, from requirement clarification to testing? If you're not seeing any movement, something is wrong.”} P1 wanted the equivalent measure against a schedule: \textit{“Maybe something like a checkpoint, like you're halfway through your milestone, or where are you at, or you're a week behind, do we need to readjust the delivery date?”} 

Across these accounts, the value participants placed on sensing was in its timing rather than in the measurement itself. P10 wanted the signal to arrive while the situation could still be corrected: \textit{“If there's early signs that could just go to the development resources first and say, hey, we're starting to turn yellow here.”}

\subsubsection{Diagnosis: Expecting AI to Combine Signals That No Single Tool Held}
\label{ex3}

Participants expected AI to go beyond detection and combine signals held in different systems into an assessment they did not think a person could assemble by hand.

P4 described this as cross referencing behavioral records with delivery outcomes: \textit{“Take some of this chat data and compare it with other things, like frequency of commits on GitHub, if things are submitted on time, NPS scores by clients.”} What P4 wanted from the combination was interpretation rather than additional metrics, because an isolated message could indicate confusion or ordinary conversation: \textit{“Trying to automate the context would be super useful.”} 

Several participants expected diagnosis to produce a score for a team or a project. P9 described a scoring practice that already existed in manual form and expected AI to take it over, noting that after each milestone an \textit{“Agile coach or Scrum Master takes that data and sort of puts a score to it manually, and then presents that to the business as the health of the team.”} P13 expected the same for completed work, asking whether a tool could \textit{“go through the project documentation or the code repo, analyze, and then give the result about the project quality.”} P2 expected the criteria themselves to be learned from the organization's project history: \textit{“You might find that there's really only like a dozen dimensions that we need to track for teamwork. And if we track those dimensions and we run a model against all the projects in the past, we see that the project has this certain percentage of success.”}

Participants also expected diagnosis to inform decisions made before a team existed. P10 expected AI to support staffing using records of prior collaboration: \textit{“If you have previous information on how those different elements of the team have worked in the past, you could probably gauge how they're going to work together as a team if they haven't before.”} 

\subsubsection{Mediation: Expecting AI to Act on Relationships Between Team Members}
\label{ex3b}

Participants also expected AI to act on the relationships between team members, which was the function they described in the most concrete terms and also the one least supported by the tools then available to them.

P11 described an intervention specified down to its trigger and its participants: \textit{“It might just be a notification that says, it's almost three days to the milestone delivery, you would want to put these three people in a meeting to have a quick huddle on Slack.”} P11 also expected AI to identify when collaborators in different time zones were actually available to one another, a coordination cost she described her teams paying daily.

Other participants expected AI to make a person's own conduct visible to them. P14 wanted feedback rather than evaluation: \textit{“Anything that kind of gives you a mirror, or see how you're being perceived even in your interactions. Because sometimes developers can be moody. So it's good to see, hey, you should watch the way you talk to people sometimes.”} P14 connected this to a gap in existing review processes, which he said assessed code but not the conduct around it: a person \textit{“can write good code, but if you don't have good personal skills, it kind of negates that sometimes.”}

Participants expected sentiment to be tracked alongside delivery. P6 wanted a project's emotional history reported together with its output: \textit{“There were 15,000 conversations, there were 10 milestones, and here's the sentiments of the team throughout.”} P6 described a light use for it, treating a positive period as a prompt to acknowledge the team rather than as a measurement. P7 expected team composition to become legible in the same way, describing an index of whether \textit{“this team has a good balance”} of interaction styles, inferred from how people wrote in Slack. P8 identified where the data for all of this would come from, and why the tools available to them did not reach it: \textit{“Multiple conversations and multiple threads often happen in isolation, meaning through DM. There's a lot of data there that is not being used for anything right now.”}

\subsection{The Limits Participants Placed on Their Team-Level Expectations}
\label{ex3c}
Participants placed three limits on what they expected: future AI should inform decisions rather than make them, whatever it collected should be disclosed to the people it collected from, and it was not the conversational assistants then becoming available, which several participants excluded directly.

The first limit governed who saw an assessment, when, and in what form. P12 described an escalation sequence that reproduced what a considerate colleague would do: \textit{“The AI model should alert the team member itself, saying, hey, I'm seeing a drop in your productivity. Let the individual see what the model is predicting, then give it time for the person to respond.”} Only if that failed would a manager be involved. P10 asked at minimum \textit{“to be notified that your manager was notified,”} and P5, who had managed teams for years, treated this as a limit on automation: \textit{“it's a very personal thing where you absolutely need to have a human involved.”} The output itself should be material for human interpretation rather than a conclusion, because participants judged the cost of a misread signal to be high. As P3 put it, \textit{“If the signal gets misinterpreted and somebody gets fired over it, that's actual harm.”} P2 described the intended output as a prompt to investigate rather than a verdict, and P14 noted that a score presented without guidance leaves its reader to draw inferences unaided.

The second limit concerned disclosure. P12 distinguished consent given at signup from consent maintained in practice: \textit{“At every given moment, people should know what is being mined and how it has been mined.”} P5 proposed aggregation as a safeguard, sharing \textit{“high level numbers, not individuals,”} with the team rather than with management alone, and P9 accepted only one of the two purposes the same instrument could serve: a measure teams could use \textit{“to know our strength, instead of to compare ourselves to another or to be capitalistic and try to demand more on a team.”}

The third limit was technological: participants excluded conversational assistants, the class of system that would soon dominate their own workflows. P2, who had used ChatGPT, judged them unsuited to the problem he had spent the interview describing, \textit{“that doesn't sound like a really effective way to go,”} allowing that they might form part of a solution without being it. He was skeptical even of the diagnostic system he had himself described, calling the search for reliable dimensions of teamwork \textit{“a massive stretch, frankly,”} and a matter of sustained research rather than a product. P8 reported the same conclusion from practice, describing the team's attempts to use ChatGPT for project planning as too generic to act on: \textit{“We couldn't take a ChatGPT response and say, here's our plan.”}

Two points carry into \autoref{sec:phase2}. The expectations recorded in 2023 were infrastructural rather than conversational, a system reading across Jira, Slack, GitHub, and project plans, learning a team's normal, and acting at a specified point within stated limits, and participants did not expect a chat interface to provide this, with some doubting that anything would. No system of that kind arrived, so the expectations were never tested against one. What arrived instead was enterprise AI in the form of personal assistants, and \autoref{sec:phase2} traces what happened to the expectations in the absence of the form they had described.

%% file: 5_phase2.tex
\section{Findings: What Became of Those Expectations in 2025 (Phase 2)}
\label{sec:phase2}

By 2025 none of the enterprise AI built around teams described in \autoref{sec:phase1} was present in participants' work. They described no system that read across Jira, Slack, and GitHub to learn a team's baseline, and none that flagged silence, scored team health, or prompted a meeting three days before a milestone. What they described instead was enterprise AI in the form of personal assistants, working at the scale of the individual task. P2 used ChatGPT and Claude for architectural and compliance documentation, and in place of Stack Overflow for \textit{“small, well-defined, but slightly non-intuitive”} programming problems. P5 built retrieval systems over company knowledge and used summarization across call transcripts and research material. P4 used it for market research, roadmap prioritization, and as a sounding board for stakeholder strategy. P6 estimated a thirty to forty percent gain in starting a task, and described the benefit as reducing the cost of beginning work rather than the cost of coordinating it. One category of tool did face the team, which was meeting transcription, and participants described it as running alongside their collaboration rather than within it. As P12 put it, \textit{“Tools like the AI note-taker automatically take notes, but we're still relying on the old methods of email or chat.”} No participant described those notes being used in later decisions.

However, the team-level expectations did not disappear with the systems that would have satisfied them. They diverged. Some participants continued to hold them, in more specific form than in 2023, and treated their absence as a matter of timing. Some remembered them and had come to consider them undesirable. Others no longer stated them at all. This divergence was specific to the team level. Participants' expectations of what AI would do for them individually rose over the same period. No participant described the outcome as a disappointment, and the requirements they had placed on AI for teamwork were by 2025 described as expectations of colleagues.

\subsection{Three Trajectories: How the 2023 Expectations of AI for Teamwork Diverged by 2025}
\label{ph0}

We group the follow-up accounts into three positions. Each concerns the team-level expectations recorded in \autoref{sec:phase1}, not participants' expectations of AI in general. \autoref{tab:trajectories} pairs each returning participant's 2023 expectation with their 2025 account. Four participants deferred the expectation, three rejected it, and three dropped it. The subsections below give the accounts in full.

\begin{table*}[ht]
\centering
\begin{threeparttable}
\caption{Team-level expectations of the ten returning participants in 2023 and their accounts in 2025. Positions are our categories, not distinctions participants drew.}
\label{tab:trajectories}
\begin{tabular}{@{}p{0.8cm}p{4.6cm}p{4.6cm}p{2.4cm}@{}}
\toprule
\textbf{ID} & \textbf{Team-level expectation in 2023} & \textbf{Account in 2025} & \textbf{Position} \\
\midrule
P5 & Compare communication volume against a team's normal; detect downtrends; share aggregate team health & Wants the synthesis he now performs by hand for the team automated & Deferred \\
\addlinespace
P6 & Detect silence; track ticket movement between lanes; report sentiment across a project & Adapt the work breakdown to how each member works while preserving traceability. Organization never settled on tooling & Deferred \\
\addlinespace
P11 & Identify cross-timezone availability; notify the three people needed before a milestone; dashboard for the engagement manager & Team feedback generated over shared documents; tools lack persistence of correction & Deferred \\
\addlinespace
P12 & Sense sentiment and urgency in Slack or Teams, opt-in, alerting the individual before the manager & A cognitive architecture for collaboration; a team assistant rather than a personal assistant; \textit{``not there yet''} & Deferred \\
\midrule
P2 & Learn dimensions predicting project success from organizational history; dashboard of check-in patterns & Cannot picture a collaborative use of AI; no ground truth about teamwork to train on & Rejected \\
\addlinespace
P10 & Predict team composition from prior collaboration; alert when a project turns yellow; improve estimation & Oversight and metrics are the wrong use; coordinating roles cannot be specified in a prompt & Rejected \\
\addlinespace
P14 & A mirror showing how one is perceived by teammates; a score of how the team is operating & Expects the capability to arrive as management instrumentation and to be used against workers & Rejected \\
\midrule
P9 & Ingest retrospectives; convert manual team-health scoring into a continuous measure; track a reassigned member over six months & \textit{``I've never had one.''} Teamwork is a domain AI has never been relevant to. Expects AI to take over programming entirely & Dropped \\
\addlinespace
P13 & Read project documentation and code repository; return an assessment of delivery quality & \textit{``AI tool has nothing to do with teamwork,''} like internet search & Dropped \\
\addlinespace
P4 & Cross-reference chat activity with commit frequency, on-time delivery, and client scores & No team-level expectation stated; expectations centered on individual strategy work & Dropped \\
\bottomrule
\end{tabular}
\end{threeparttable}
\end{table*}

\subsubsection{Deferred: Keeping the Expectation and Making It More Specific}
\label{ph4}

Four participants still held their 2023 team-level expectations in 2025 and described them in more detail than before. Their expectations were deferred rather than met: the systems had not arrived, and these participants treated that as a matter of timing. What distinguishes this group is not that they remained optimistic but the reasons they gave for the absence. None concluded that the expectation was infeasible or undesirable.

For P12, available tools were built for the wrong user: the individual rather than the team. His forecast, he said, had not been met and had not been withdrawn: \textit{“Some of my predictions have not yet come true, but I still think it will still come.”} What he specified in 2025 went beyond the sentiment sensing he had described in 2023: \textit{“If somebody comes up with the right cognitive architecture about collaboration using AI tools, that would be really transformational. Not just the emotions, but also intent, understanding the intent, and even actually helping the team align their action towards the core objective or the goals of your project.”} He put the difference as one of unit rather than capability: \textit{“Think about a team assistant, not a usual personal assistant.”} On the state of available tools he was direct: \textit{“Collaboration is one big area where AI can help a lot. And it is not there yet.”}

P6 had adopted none of the tooling he wanted in 2023, and explained this as his organization never having settled on any. His expectation had meanwhile shifted from detecting trouble to allocating work: \textit{“If a project has a structure, here's the Jira board, here's how the user stories break, here's the work breakdown structure. What would be really cool is having AI adapt it to how people work best.”} The example came from his own team, where some members needed only key tasks and brief comments while others worked from detailed checklists. The difficulty he identified was holding both together: any such adaptation had to preserve \textit{“traceability, metrics, visibility”} across the project as a whole.

What P11 now wanted ran over shared documents rather than communication logs: \textit{“If a team collaborates in the same AI tool, it is then able to give performance feedback. If you do a brainstorming session and everybody has contributed to that document, it could then at the end of the week give personalized feedback and say, at 5pm on Tuesday we had this conversation, you added this to the document, this is what would be the next step for you.”} The obstacle P11 named was not capability but persistence. A person given feedback returns having acted on it, and the tools in use did not, so the same corrections had to be supplied again each time.

P5 was already doing this work by hand. Asked what he wanted from AI for his team, he described his own current responsibilities: \textit{“Even though we do have transcription and summary services, I do some stuff to augment that even further and share that back to the team. I've been the one that's been ingesting all the data and presenting it in a format that actually makes sense to get everybody else on the same page. I would love to automate that even further. That's one huge area for improving team collaboration.”} On P5's team the integration function expected of AI in 2023 was, by his own account, being performed by him.

Between 2023 and 2025 these expectations became more specific rather than less, and the obstacles participants named were organizational and architectural: the wrong unit of design, tooling never adopted, no memory of correction, and work not yet automated.

\subsubsection{Rejected: Remembering the Expectation and Turning Against It}
\label{ph3}

Three participants recalled what they had wanted and by 2025 considered it unattainable, misdirected, or likely to arrive in a form they did not want.

In 2023 P10 had wanted prediction of team composition from records of prior collaboration, alerts when a project began to \textit{“turn yellow,”} and estimates accurate enough to relieve developers of guessing. In 2025 he rejected that use: \textit{“I'm not convinced that more of those is a good thing. Some of them are good to have, it's always nice to have a pulse on things. But using it as a development tool versus trying to give oversight and metrics, I think, is the better approach, and encouraging teammates to do that.”} His reason concerned what such a system would have to represent. The coordinating roles he had once expected AI to inform could not be specified, because, in his account, the knowledge they depend on is not recorded anywhere: \textit{“It's not as easy as putting a system prompt in there to say you are a product owner or you are a project manager. It's just not that sterile of an environment. Everybody's job is very organic. You do things that only you know about the company, and you can't system prompt your way out of it.”}

P10 still wanted the underlying problem addressed, but placed it beyond what he saw current tools reaching: \textit{“I've seen the value of the interpersonal experience. That value really got lost once we went remote, and I don't know that there's an AI tool that solves for a lot of that. But if you can think of one, I think you'd have a billion dollar industry there.”}

P2's position in 2025 extended skepticism he had already voiced in 2023, when he called the search for measurable dimensions of teamwork \textit{“a massive stretch.”} By 2025 he said he could not picture what a collaborative use of AI would consist of: \textit{“There isn't any direct collaboration with AI tools, rather people use them as an assistant, in the same way that you'd use a search engine. I can't really even visualize how you would collaborate with other people within AI.”} He explained this in terms of what such a system would have to be trained on, arguing that there is \textit{“no concrete reality you could put into a database”} for good teamwork, because the judgment is always made by \textit{“some individual human being.”} He allowed that such a system could exist without seeing how it would be reached: \textit{“The best AI tool would mimic that behavior. I can imagine something like that existing. I don't see the bridge between where we are and there.”}

P14's rejection took a different form. In 2023 he had wanted a system that would show him how he came across to teammates, feedback rather than evaluation. In 2025 he still expected the capability to arrive, but as management instrumentation used against workers, citing a widely circulated case of a company tracking productivity by camera and observing that \textit{“instead of using the feedback to help them, you're just going to berate them or find ways to cut them.”} He identified the incentive directly: \textit{``Why do I have to pay somebody with benefits and a salary when I can just get some GPUs?''} P2 and P10 no longer believed the system could be built. P14 believed it would be, and what he rejected was not the mirror he had wanted but the form he now expected it to take: the same capability, held by management rather than by the person it described. 

\subsubsection{Dropped: No Longer Stating the Expectation}
\label{ph2}

Three participants no longer stated their 2023 team-level expectation in 2025. Two of them, P9 and P13, described teamwork as a domain AI had never been relevant to. The third, P4, described expectations centered on his own strategy work and did not return to the cross-referencing system he had specified in 2023. All three continued to hold substantial expectations of AI in other domains. We cannot determine from interview data whether they had forgotten their 2023 accounts, no longer endorsed them, or did not treat them as relevant to the question asked. What we can observe is that the expectation is absent from their 2025 accounts, and that P9 and P13 offered reasons why such an expectation would not make sense.

In 2023, P9 had described how AI should ingest retrospective meetings, convert what an Agile coach scored by hand into a continuous measure of team health, and track a contributor moved to a new team over the following six months. In 2025, asked what he had expected AI to do for his team, P9 said: \textit{“I didn't have expectation of AI for teamwork. My mind isn't thinking about a particular application that would impact my teamwork. I've never had one.”} He explained why he considered the category inapplicable: \textit{“Human wants to talk to humans, human wants to stay together with human, human wants to fellowship with human. And that team bonding, I don't really know whether AI would really be able to succeed in taking out that space.”} Asked again later in the interview, he drew the same boundary: \textit{“On a personal level, AI has definitely impacted that. But on a team level, I don't have any expectations for it much.”}

P9's expectations of AI had not diminished in general. In the same interview he expected AI to handle most business calls, to negotiate with other AI systems on his behalf, and eventually to write software without him. What he no longer applied those expectations to was teamwork.

P13 arrived at the same position from a different starting point. In 2023 he had expected AI to read a project's documentation and code repository and return an assessment of delivery quality. In 2025 he compared AI to a tool no one evaluates on collaborative terms: \textit{“AI will make individuals more self-competent. Just like internet browser search has nothing to do with teamwork, AI tool has nothing to do with teamwork. Productivity in the project can improve overall.”}

The comparison does what P9's account does. Search is not a technology that failed at collaboration; it is one that was never a candidate. In that framing the 2023 expectation is not merely unmet. It has dropped out of what P9 and P13 would think to ask for. P4 gave no such reason; the expectation is simply no longer in his account.

\subsection{From Team-Level AI Expectations to Colleague Norms}
\label{ph5}

The three trajectories describe what became of participants' earlier expectations for AI in teamwork. Across them, however, participants shared another feature in their 2025 accounts: none described the absence of the team-level systems they had imagined as a loss. Instead, they evaluated AI primarily through what it now did for them individually, while some concerns that had appeared in their earlier expectations for AI in teamwork were articulated as expectations of colleagues.

Participants across all three trajectories described personal AI as having exceeded their expectations. P2, who by 2025 could not picture a collaborative use of AI, described the preceding two years as a series of surprises: `My expectations, whatever they were, were significantly exceeded.'' P5 similarly recalled the impact of his first week using generative AI, while P9 expected AI eventually to take over programming altogether. Yet participants did not extend these assessments to teamwork. Later in the same interview, P2 said that `in terms of teamwork, I wouldn't say it really had a major impact.'' The two judgments were not presented as contradictory. The benefits of personal AI were immediate and visible in specific tasks, whereas the team-level systems participants had imagined in 2023 had never entered their work.

At the same time, several concerns that participants had once associated with AI for teamwork appeared in 2025 as expectations of other people. Pace provides one example. P9 explained that AI had shortened the time needed to complete programming work and, as a result, `expectation is now very high from managers and team members.'' Pace had already been a source of coordination friction in 2023, but by 2025 participants explicitly connected heightened expectations of one another to the speed enabled by personal AI. Disclosure showed a similar pattern. In 2023, participants had emphasized that systems collecting information about workers should make that collection visible. By 2025, disclosure was instead discussed as a norm governing colleagues' use of AI. As P11 put it, using AI was acceptable; concealing that use was not: `It would only be an issue if they didn't tell anyone they were doing it.''

We do not interpret this as evidence that requirements literally transferred from AI systems to colleagues. Rather, similar concerns about pace, visibility, and accountability became articulated through social expectations once the anticipated team-level systems were absent. This pattern is secondary to the three trajectories above, but it further illustrates how the surrounding design space changed between 2023 and 2025. As AI capabilities fell short of what participants had initially expected of them, what they once discussed partly as a problem for future AI systems was increasingly discussed as a problem of how people should work with one another in an AI-mediated workplace.

%% file: 6_discussion.tex
\section{Discussion}
\label{sec:discussion}

We followed practitioners' expectations for AI in teamwork across a period in which generative AI moved rapidly from novelty to everyday work. By 2025, expectations articulated in 2023 had not simply strengthened or weakened; they had acquired different histories. Some remained open, some were reconsidered and rejected, and some disappeared from participants' later accounts. We argue that this matters for HCI because the design space visible in what users say today is not ahistorical. Longitudinal records reveal not only \emph{what} users currently articulate, but what has persisted, what has become objectionable, and what has fallen out of view.

\subsection{Expectations Have Histories, Not Just Current States}
\label{dis1}

Most user research gives designers a state: what someone wants, needs, or finds acceptable at the moment they are asked. Our two phases reveal something different---the history behind that state. A participant who did not ask for a particular form of AI in 2025 might have rejected an expectation they previously held, might no longer articulate a possibility documented in 2023, or might never have considered it at all. These positions can look similar in a cross-sectional study but carry very different meanings.

The dropped expectations make this clearest. P9 and P13 described concrete possibilities for AI in teamwork in 2023, yet in 2025 described teamwork as a domain to which AI was not relevant. The earlier interviews do not prove that an underlying need persisted: those expectations may have been forgotten, abandoned, or no longer considered relevant. What they establish is that a possibility once available in participants' accounts was no longer recoverable from the later account alone.

HCI has long developed methods for helping people articulate possibilities that are difficult to imagine~\cite{norman2013design,patnaik1999needfinding}, including participatory design~\cite{bannon2012design,steen2013co,yu2025participatory}, probes~\cite{boehner2007hci,ibrahim2024systematic}, and design fiction~\cite{rapp2019design,sengers2021speculation}. Our findings expose the inverse problem: a possibility can be articulable at one moment and become unavailable in a later account. The challenge is therefore not only how to elicit possibilities, but how to preserve their histories once they have been articulated.

\subsection{The Design Space Users Articulate Is Historically Contingent}
\label{dis2}

The divergence occurred while participants' experience with AI was changing substantially. In 2023, generative AI was only beginning to enter their work. By 2025, personal assistants were routine across many individual tasks, while the systems participants had imagined for sensing, diagnosing, and mediating teamwork had not entered their work. Research on technological expectations similarly argues that what people anticipate develops in relation to what has been built, experienced, and made salient around them~\cite{borup2006sociology,vanlente2012navigating}.

Our data do not establish that personal assistants caused participants to defer, reject, or drop earlier expectations. They show something narrower but important for HCI: the same question was answered from different technological worlds. Expectations are therefore not observations made from outside technological change; they are articulated from within it. Neither the 2023 nor the 2025 account is ground truth. Each makes some possibilities salient while leaving others less visible.

This temporal contingency matters because HCI often uses present-day accounts to orient future design. A requirement gathered today may be well grounded in today's experience and still omit alternatives that were salient earlier. Longitudinal comparison makes those absences visible and turns the history of an expectation into evidence rather than noise.

\subsection{Different Expectation Trajectories Offer Different Design Evidence}
\label{dis3}

The main design implication of our findings is not to simply revisit users more often, but to capture and make sense of the \emph{trajectory} of users' expectations as part of the design evidence. Deferred, rejected, and dropped expectations each tell designers something different.

\paragraph{\textit{\textbf{Deferred expectations identify unresolved opportunities.}}}
A deferred expectation remains desirable to the participant even though the corresponding system has not materialized in their work. The value of the second interview is not merely confirmation that the expectation persists. Participants often made their earlier ideas more specific, identifying requirements around persistence, adaptation, integration, and adoption that were less visible in 2023. As enterprise AI gains new capabilities, including agentic systems and broader access to workplace context~\cite{shao2025future,Spataro2025FrontierFirm}, these expectations mark possibilities whose feasibility may have changed while their motivating problems remained salient to participants. Their trajectory tells designers what is worth revisiting and what later experience has added to the original specification.

\paragraph{\textit{\textbf{Rejected expectations produce negative design knowledge.}}}
Rejection is not simply the disappearance of demand, that is, participants no longer wanting the system they had described before. It reveals why an earlier possibility became unacceptable. P2 and P10 questioned whether ``good teamwork'' could be represented reliably enough for AI to score or predict it; P14 worried that supportive feedback could become managerial surveillance. These objections identify boundary conditions for design. They suggest separating information assembly from automated judgment, preserving human interpretation where no credible ground truth exists, and treating access and escalation as governance questions rather than implementation details. An expectation that has been rejected can therefore remain productive for design because the reason for rejection shows what a redesigned system would have to overcome.

\paragraph{\textit{\textbf{Dropped expectations preserve alternatives that current elicitation cannot surface.}}}
Dropped expectations are different because the later account no longer contains the possibility at all. Here the earlier record acts as an archive of an alternative design direction. Designers can deliberately reintroduce that alternative through a scenario, probe, mock-up, or lightweight prototype and ask how it is received now. The purpose is not to recover a supposedly ``true'' earlier need. It is to distinguish among possibilities that are now clearly unwanted, possibilities that have become irrelevant, and possibilities that ceased to be salient without ever being tested. Design fiction and related methods already make possible futures concrete enough to discuss~\cite{rapp2019design,sengers2021speculation}; longitudinal records can tell us which lost alternatives to put back on the table.

These three trajectories imply a different role for longitudinal research in design. Its value is not that repeated elicitation converges on an increasingly accurate requirements document. It preserves \emph{differences across moments}. Prior longitudinal qualitative work in HCI and CSCW demonstrates how returning to the same participants reveals change unavailable in cross-sectional studies~\cite{blaising2021making,jungselius2025tracing}. Our findings suggest that these differences can themselves be actionable: deferred expectations identify opportunities to revisit; rejected expectations reveal constraints; dropped expectations preserve alternatives that may otherwise disappear from consideration.

This approach complements rapid iteration rather than competing with it~\cite{reis2011lean,davidoff2007rapidly,hussain2009current}. Prototyping and testing can tell designers whether a current concept works for current users. They cannot reveal that another possibility was salient two years earlier if no record of it remains. Longitudinal elicitation contributes a different form of evidence: a history of how the design space changed.

The approach also has limits. Our two phases reveal divergence over two years in one organization; they do not establish how quickly expectations evolve elsewhere or how frequently researchers should return. Nor does a trajectory tell designers which expectation should ultimately be acted upon. Repeated qualitative work carries costs, although AI-assisted transcription and analysis may reduce some of the burden~\cite{aubin2024llms,chen2025echomind}. We therefore do not argue for longitudinal panels in every design project. The narrower point is that in fast-moving technological domains, earlier accounts can retain design information that newer accounts do not supersede.

\subsection{Designing "AI for Teamwork" From Understanding Expectation Trajectories}
\label{design}

The three trajectories suggest three different design responses. Deferred expectations can be refined using what participants learned later; rejected expectations can be redesigned around the reasons for rejection; and dropped expectations can be deliberately reintroduced to test whether they still matter. We use our own case to show what this looks like in practice: for each trajectory, we take the team-level expectations our participants articulated in 2023 and the accounts they gave in 2025, and develop one design concept for AI in teamwork from them. The concepts are illustrations of how a trajectory can inform design, not a design agenda for this organization. Recent advances in agents and systems that can access information across workplace tools make several of these possibilities easier to instantiate than they were in 2023~\cite{shao2025future,Moveworks2025AIAgentMarketplace,Spataro2025FrontierFirm}. The goal is not to argue that these three systems should be built, but to show, through one worked example, how the trajectory of an expectation changes what a designer does with it.

\begin{table*}[ht]
\centering
\caption{Three design responses to different expectation trajectories.}
\label{tab:design-implications}
\begin{tabular}{p{2.2cm}p{3.0cm}p{3.2cm}p{4.3cm}}
\hline
\textbf{Trajectory} &
\textbf{Design response} &
\textbf{What changes} &
\textbf{Illustrative concept} \\
\hline

\textbf{Deferred} &
\textbf{Refine} what remains desirable &
Use later experience to make the earlier idea more specific and easier to adopt &
\textbf{Project Memory}: maintain persistent context across tickets, conversations, documents, and project history \\

\textbf{Rejected} &
\textbf{Redesign} around objections &
Preserve useful functions while removing or constraining what participants came to reject &
\textbf{Evidence Lens}: connect evidence across tools without scoring the team or making personnel judgments \\

\textbf{Dropped} &
\textbf{Reintroduce} what disappeared &
Make the earlier possibility concrete again and learn from current responses &
\textbf{Reconnect Prompt}: offer a lightweight coordination intervention and observe whether users accept, reject, or ignore it \\

\hline
\end{tabular}
\end{table*}

\paragraph{\textit{\textbf{Deferred $\rightarrow$ Refine: Project Memory.}}}
Deferred expectations remained meaningful to participants, but later accounts often made clearer what those systems would need to do. In 2023, participants in our study imagined AI that could understand team activity across tools. By 2025, those who retained this expectation emphasized needs that were less explicit earlier: persistence across interactions, shared context across people, and continuity across project activity. P5 was already performing some of this integration manually, while P11 and P12 described systems that could retain context beyond a single interaction or individual user.

\textit{Project Memory} builds on this trajectory. Rather than treating each AI interaction as separate, it would maintain persistent context across tickets, conversations, documents, meetings, and project history. The point is not simply to implement the 2023 expectation. The deferred trajectory tells us how to refine it: the earlier possibility remains, while later experience clarifies the importance of persistence, shared context, and fitting into existing coordination practices.

\paragraph{\textit{\textbf{Rejected $\rightarrow$ Redesign: Evidence Lens.}}}
Rejected expectations provide a different kind of guidance. In 2023, participants imagined AI combining signals across Jira, Slack, GitHub, and other sources to assess project or team health. By 2025, some participants rejected the assumption that AI could reliably determine what ``good teamwork'' looks like, while others worried that such evaluation could become managerial surveillance. The later objection therefore targets particular parts of the earlier design, not necessarily the underlying problem of fragmented information.

\textit{Evidence Lens} preserves the information-integration function while removing the automated verdict. A user could ask a bounded question such as ``why is this stalled?'' and receive the relevant ticket history, discussion threads, commits, and meeting records, with their sources visible. The system would not assign a team-health score or make a personnel judgment. Here, rejection tells the designer where to draw a boundary: keep the synthesis participants valued, but redesign around concerns about validity, interpretation, and oversight.

\paragraph{\textit{\textbf{Dropped $\rightarrow$ Reintroduce: Reconnect Prompt.}}}
Dropped expectations require a third response because the later account no longer contains the possibility at all for some people. In 2023, participants P9, P13, and P4 had each wanted AI to read a team's own records and tell the team how it was doing, imagining AI that would notice when coordination was breaking down and help reconnect the right people. By 2025, none of the three stated a team-level expectation, and the reasons P9 and P13 gave placed AI on the other side of a line: teamwork was human bonding, and AI, like search, made individuals more capable and had \textit{``nothing to do with teamwork.''} The longitudinal record does not tell us whether the earlier idea remains desirable, but it preserves an alternative that present-day elicitation would not surface.

\textit{Reconnect Prompt} reintroduces that possibility in the smallest form that stays on the team side of the line these participants drew. Before an upcoming milestone, the system could identify collaborators whose interdependent work has seen little recent interaction and ask whether they should reconnect. The prompt is not meant to assume that AI should coordinate the team. It functions as a probe: acceptance, rejection, or indifference from P9, P13, and P4 provides new evidence about whether their 2025 position was that AI has no place in teamwork at all, or only that it should not judge a team, a distinction their accounts do not settle.  This use parallels design fiction and related methods that make possible technologies concrete enough for participants to respond to~\cite{rapp2019design,sengers2021speculation}.

Together, these examples show how expectation trajectories can change the design response. Deferred expectations suggest what to refine; rejected expectations reveal what must be redesigned; and dropped expectations identify possibilities that may need to be reintroduced before their current relevance can be assessed. The value of the longitudinal record is therefore not that it tells designers which moment was ``right,'' but that it shows what happened to a design possibility and makes that history usable in design.

%% file: 7.limitations.tex
\section{Limitations and Future Work}
\label{limitations}

Our study has several limitations that bound how our findings should be read. First, the sample is relatively small and the trajectory groups are smaller. Fifteen participants took part in Phase 1 and ten in Phase 2, and the three positions we describe rest on cells of four, three, and three. The scale is in line with two-phase interview studies in HCI and CSCW~\cite{blaising2021making,jungselius2025tracing}, and our claims are sized to it: we show that these trajectories occur and how they cohere, not how they are distributed. In particular, the three cases of a dropped expectation establish that the pattern exists, not how often it occurs. Establishing prevalence, including how often an expectation is dropped rather than deferred or rejected, would require surveys or archival analyses of how workers' stated expectations of AI have changed over the same period.

Second, this is a single-organization case study. All participants worked in one distributed, project-based software development organization that adopted AI early. We chose depth within one case over breadth across several, since the question we ask, what became of expectations formed as the technology arrived, requires following the same people in the same setting across that arrival. The choice bounds transferability, and a case study of this kind supports claims about what can happen and how, not about how often. Teamwork in this organization is modular and contract-based, and its members are disposed to experiment. Expectations of enterprise AI for teamwork may follow different paths in co-located teams, in sectors such as healthcare or finance where deployment is constrained, or in organizations that adopted AI later and formed their expectations after personal assistants were already the established form. 

Third, our data are interview accounts, not unobstructive behavioral data. We observe what participants said they expected and said became of it, not what they privately believed or how their teams actually worked, and the interview itself is part of the setting: asking people to revisit their earlier statements may prompt reconciliations they would not otherwise produce. We have marked the point where this limit matters most. A dropped expectation, as we observe it, is a fact about what participants said, and interview methods cannot separate it from forgetting or quiet disavowal. Combining repeat interviews with communication logs or observation would show whether the relocation we describe, of pace and disclosure into peer norms, is visible in how teams actually interact.

Fourth, two time points bracket the change but do not show its course. We cannot say when between 2023 and 2025 each expectation was deferred, rejected, or dropped, or what occasioned it, and we cannot see what happens next. Products marketed as team assistants have begun to appear since our second phase, and the four participants who deferred their expectations suggest the idea could become sayable again once something makes it concrete. A third phase with the same participants would show whether dropped expectations stay dropped, whether reopened expectations resemble the 2023 ones, and whether the requirements now carried by colleagues return to systems when systems appear to carry them. That is the longer record \autoref{dis3} calls for, and this study is its first two points.

%% file: 8_conclusion.tex
\section{Conclusion}
\label{sec:conclusion}

We asked members of a distributed, project-based software development organization what they expected AI to do for their teams, first in 2023 and again in 2025. In 2023 they expected enterprise AI built around teams. By 2025 no such system had reached them and personal AI assistants had become routine. The same expectation, held by the same people, followed three trajectories: some deferred it, some rejected it, and some no longer stated it. No one described this as a disappointment, and the requirements once addressed to AI were applied by colleagues to one another instead.

For HCI, the implication is that a present-day user account conceals its own history. A participant who does not ask for a system today may still want it, may have turned against it, or may no longer think to ask, and a single round of elicitation cannot tell these apart. Each of these trajectories is different design material: is different design material: deferred expectations mark opportunities to refine, rejected expectations mark constraints to redesign around, and dropped expectations preserve alternatives that an earlier record makes it possible to reintroduce. In rapidly evolving domains, earlier accounts are not expired results but evidence of how the design space users articulate has changed.